\documentclass{PoS}
\usepackage{epsfig}

\title{The Giant Radio Array for Neutrino Detection (GRAND): Present and Perspectives}

\ShortTitle{Giant Radio Array for Neutrino Detection}

 \author{\speaker{Ke Fang}$^{1}$\footnote{  E-mail: {kefang@umd.edu}},
  Jaime \'Alvarez-Mu\~niz$^{2}$,
 Rafael Alves Batista$^3$,
  Mauricio Bustamante$^{4}$, 
  Washington Carvalho$^5$, Didier Charrier$^6$, Isma\"el Cognard$^{7,8}$, Sijbrand De Jong$^9$, Krijn D. de Vries$^{10}$, Chad Finley$^{11}$, Quanbu Gou$^{12}$, Junhua Gu$^{13}$, Claire Gu\'epin$^{14}$,  Jordan Hanson$^4$, Hongbo Hu$^{12}$, Kumiko Kotera$^{14}$, Sandra Le Coz$^{13}$, Yi Mao$^{15}$, Olivier Martineau-Huynh$^{16}$, Clementina Medina$^{16, 17}$, Miguel Mostafa$^{18}$, Fabrice Mottez$^{19}$, Kohta Murase$^{18}$, Valentin Niess$^{19}$, Foteini Oikonomou$^{18}$, Frank Schr\"oder$^{20}$, Cyril Tasse$^{21}$, Charles Timmermans$^9$, Nicolas Renault-Tinacci$^{14}$, Mat\'ias Tueros$^{17}$, Xiang-Ping Wu$^{13}$, Philippe Zarka$^{22}$, Andreas Zech$^{19}$, Yi Zhang$^{12}$ , Qian Zheng$^{13}$, Anne Zilles$^{14}$ \\ 
     \begin{changemargin}{0cm}{-0.5cm}  
\begin{flushleft}
      {     {$^{1}$}Department of Astronomy, Joint Space-Science Institute, University of Maryland, College Park, MD, 20742, USA\\
      $^{2}$}Departamento de F\'isica de Part\'iculas,  Universidade de Santiago de Compostela, 15782, Santiago de Compostela, Spain\\
{$^{3}$}Department of Physics - Astrophysics, University of Oxford, DWB, Keble Road, OX1 3RH, Oxford,
UK\\
   {$^4$}Center for Cosmology and AstroParticle Physics, Department of Physics, Ohio State University, Columbus, OH 43210, USA\\
  {$^5$}Physics institute, University of São Paulo, Rua do Matão, trav. R, Cid. Universitária, São Paulo, Brazil \\
   {$^6$}SUBATECH, CNRS-IN2P3, Université de Nantes, Ecole des Mines de Nantes, Nantes, France\\
{$^7$}Laboratoire de Physique et Chimie de l'Environnement et de l'Espace LPC2E CNRS-Universit\'e d'Orl\'eans, F-45071 Orl\'eans, France\\
      {$^8$}Station de radioastronomie de Nan\c{c}ay, Observatoire de Paris, CNRS/INSU F-18330 Nan\c{c}ay, France\\
    {$^9$}Nikhef/Radboud University, Nijmegen, the Netherlands\\
  {$^{10}$}Vrije Universiteit Brussel, Dienst ELEM, B-1050 Brussels, Belgium\\
    {$^{11}$}Oskar Klein Centre and Dept. of Physics, Stockholm University, SE-10691 Stockholm, Sweden\\
   {$^{12}$}Key Lab of Particle Astrophysics, IHEP, Chinese Academy of Sciences, Beijing 100049, China\\
    {$^{13}$}National Astronomical Observatory, Chinese Academy of Sciences, Beijing 100012, China\\
 {$^{14}$}Sorbonne Univ., Paris 6, CNRS UMR 7095, Institut d'Astrophysique de Paris, 98 bis bd Arago, 75014 Paris, France\\
   {$^{15}$}Department of Physics, Tsinghua Center for Astrophysics, Tsinghua University, Beijing 100084, China\\
     {$^{16}$}LPNHE, CNRS-IN2P3 and Universités Paris 6 \& 7, BP200, 4 place Jussieu, 75252 Paris, France\\
     {$^{17}$}Instituto de Física La Plata, CONICET CCT-La Plata, Calle 49 esquina 115 (1900), La Plata, Argentina\\
   {$^{18}$}Dept. of Physics, Dept. of Astronomy \& Astrophysics, Penn State University, University Park, PA, USA\\
      {$^{19}$}Clermont Universit\'e, Universit\'e Blaise Pascal, CNRS/IN2P3, Laboratoire de Physique Corpusculaire, BP.\ 10448,  63000 Clermond-Ferrand, France\\
   {$^{20}$}Institut f\"ur Kernphysik, Karlsruhe Institute of Technology (KIT), Karlsruhe, Germany\\
   {$^{21}$}LUTH, Observatoire de Paris, CNRS, Université Paris Diderot, PSL Research University, 5 place Jules Janssen, 92190, Meudon, France\\
 {$^{22}$}LESIA \& USN, Observatoire de Paris, CNRS, PSL, UPMC-SU/UPD/UO, 92195 Meudon, France       
\end{flushleft}
         \end{changemargin}
     }

\abstract{The Giant Radio Array for Neutrino Detection (GRAND) aims at detecting ultra-high energy extraterrestrial neutrinos via the extensive air showers induced by the decay of tau leptons created in the interaction of neutrinos under the Earth's surface. Consisting of an array of $\sim10^5$ radio antennas deployed over $\sim 2\times10^5\,\rm {km}^2$, GRAND plans to reach, for the first time, an all-flavor sensitivity of $\sim1.5\times10^{-10} \,\rm GeV\, cm^{-2} \,s^{-1}\, sr^{-1}$ above $5\times10^{17}$~eV and a sub-degree angular resolution, beyond the reach of other planned detectors. We describe here preliminary designs and simulation results, plans for the ongoing, staged approach to the construction of GRAND, and the rich research program made possible by GRAND's design sensitivity and angular resolution.
}

\FullConference{35th International Cosmic Ray Conference --- ICRC2017\\
		10--20 July, 2017\\
		Bexco, Busan, Korea}

\begin{document}

\section{Introduction}
% why do we need GRAND, in addition to existing in-ice detectors
The workings of the nature's highest-energy accelerators are a half-century-old mystery \cite{KO11}. It is difficult to use ultra-high energy cosmic rays (UHECR, with $E>10^{18}$~eV) to pinpoint a source as these charged particles can be deflected by up to tens of degrees by extragalactic and Galactic magnetic fields. On the contrary, secondary neutrinos, produced when cosmic rays interact with matter inside a source or with the cosmic microwave background (CMB) during intergalactic propagation, serve as direct probes, as neutrinos are not deflected and barely interact over cosmological distances. 
Two steps need to be achieved toward finding sources of UHE neutrinos:
\begin{enumerate}
\item{Detecting an UHE neutrino. Depending on the chemical composition and maximum energy of UHECRs, the flux of cosmogenic neutrino background is expected at a level of $\sim10^{-9}-10^{-8}\,\rm GeV\,cm^{-2}\,s^{-1}\,sr^{-1}$ per flavor at EeV energies. A small number of diffuse neutrinos found in this step, however, will not reveal an UHE neutrino point source. }
\item{Identifying individual sources. This step requires the detection of a sizable number of neutrinos to allow point sources to stand out of a diffuse background ~\cite{Ahlers:2014ioa, FKMMO16}. For instance, if sources follow a uniform number density of $10^{-7}\,\rm Mpc^{-3}$, a $5\sigma$ detection of point sources requires measuring $\sim 200$ neutrino events with a sub-degree angular resolution~\cite{FKMMO16}.  }
\end{enumerate}

Astrophysical tau neutrinos ($\nu_\tau$) can be detected through extensive air showers (EAS) induced by tau ($\tau$) decays in the atmosphere  \cite{1999ICRC....2..396F, 2002APh....17..183B}. Specifically, a cosmic $\nu_\tau$ can produce tau leptons under the Earth surface via charge-current interactions. The short-lived $\tau$ decays in the atmosphere after traveling tens of kilometers and generates an EAS that emits coherent electromagnetic emissions up to frequencies of hundreds of MHz.  Such inclined showers have been measured and confirmed by the Auger Engineering Radio Array to have a radio footprint covering several km$^2$ \cite{refId0}. 
With practical advantages such as low unit cost, easiness of deployment and maintenance, radio antennas present appealing characteristics for the detection of very inclined showers \cite{2017PrPNP..93....1S}.

Consisting of $\sim10^5$ radio antennas covering $200,000\,\rm km^2$ in a mountainous area, the Giant Radio Array for Neutrino Detection (GRAND) will detect air showers generated by Earth-skimming UHE tau neutrinos with unprecedented sensitivity. Mountain ranges with low electromagnetic background noise are preferable sites, as mountains provide both additional interaction targets and a slanted surface that is more favorable to detect inclined showers compared to flat ground.  
The antennas are foreseen to operate in the 60-200~MHz band to avoid the short-wave background at lower frequencies. The highest frequencies enable detection of the Cherenkov ring associated with the air shower \cite{2012PhRvD..86l3007A}, which can be used as an unambiguous signature to reject background.

\section{Detector performance}

\subsection{Sensitivity}

\begin{figure}[h]
\centering
\epsfig{file=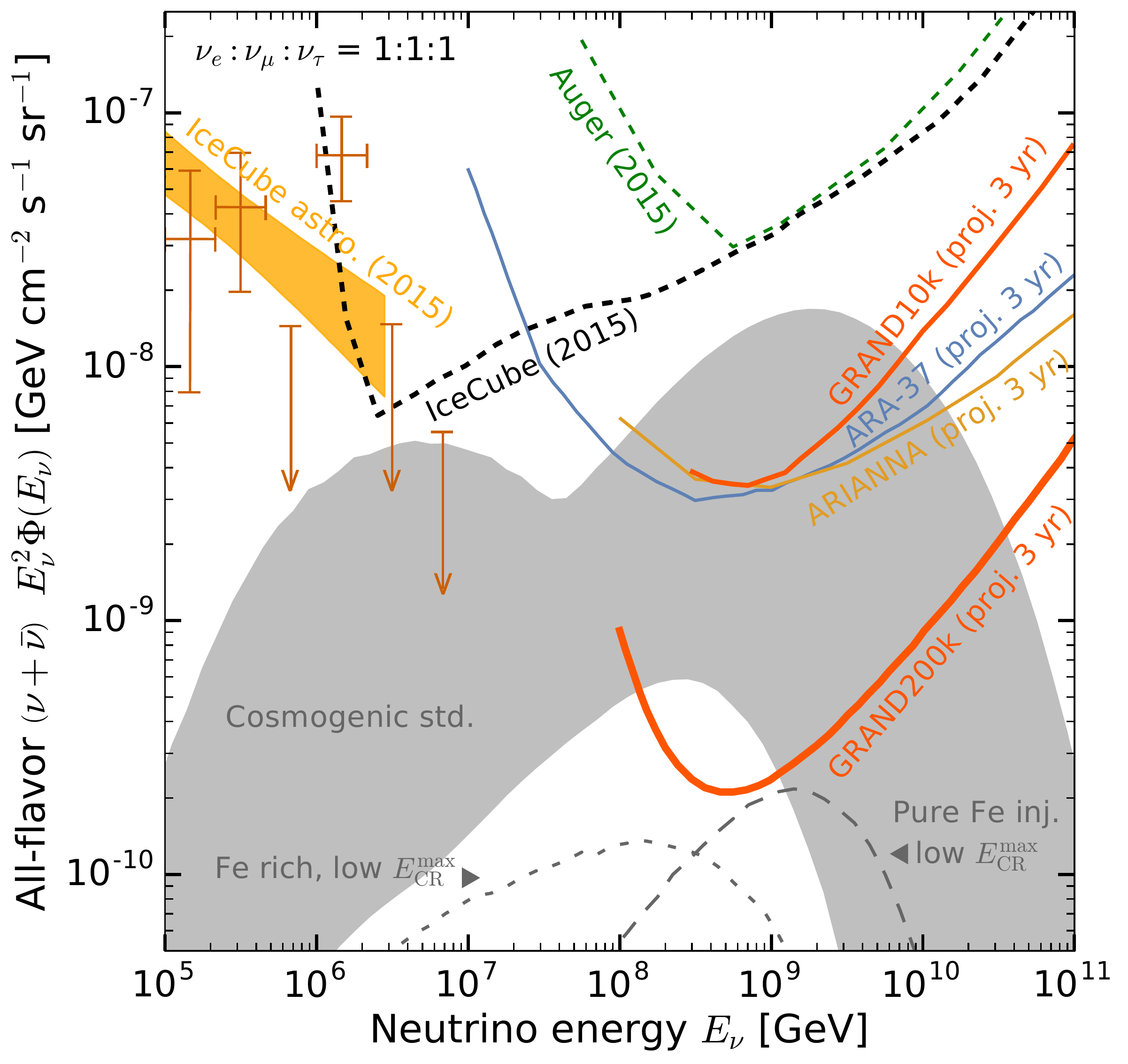,width=0.65\textwidth}  
\caption{\label{fig:sensitivity}
{\small Three-year differential sensitivity of a 10,000 $\rm km^2$ simulated setup (red thin curve) and of the projected GRAND array (red thick curve). The grey shaded region brackets fluxes of cosmogenic neutrinos from models with different cosmic ray compositions, spectra, and source evolution histories that provide reasonable fits to the Auger UHECR spectrum \cite{2010JCAP...10..013K}. The grey dashed and dotted lines indicate the predicted cosmogenic neutrino flux in a pessimistic scenario where the bulk of UHECRs are iron nuclei. High-energy neutrinos measured by IceCube \cite{2015arXiv151005223T}, UHE neutrino sensitivities of IceCube \cite{2016PhRvL.117x1101A}, Auger \cite{PhysRevD.91.092008}, ARA \cite{2012APh....35..457A} and ARIANA \cite{2015APh....70...12B} are also shown for reference. }}
\end{figure}

\begin{figure}[h]
\vspace*{-6mm}
\centering
\epsfig{file=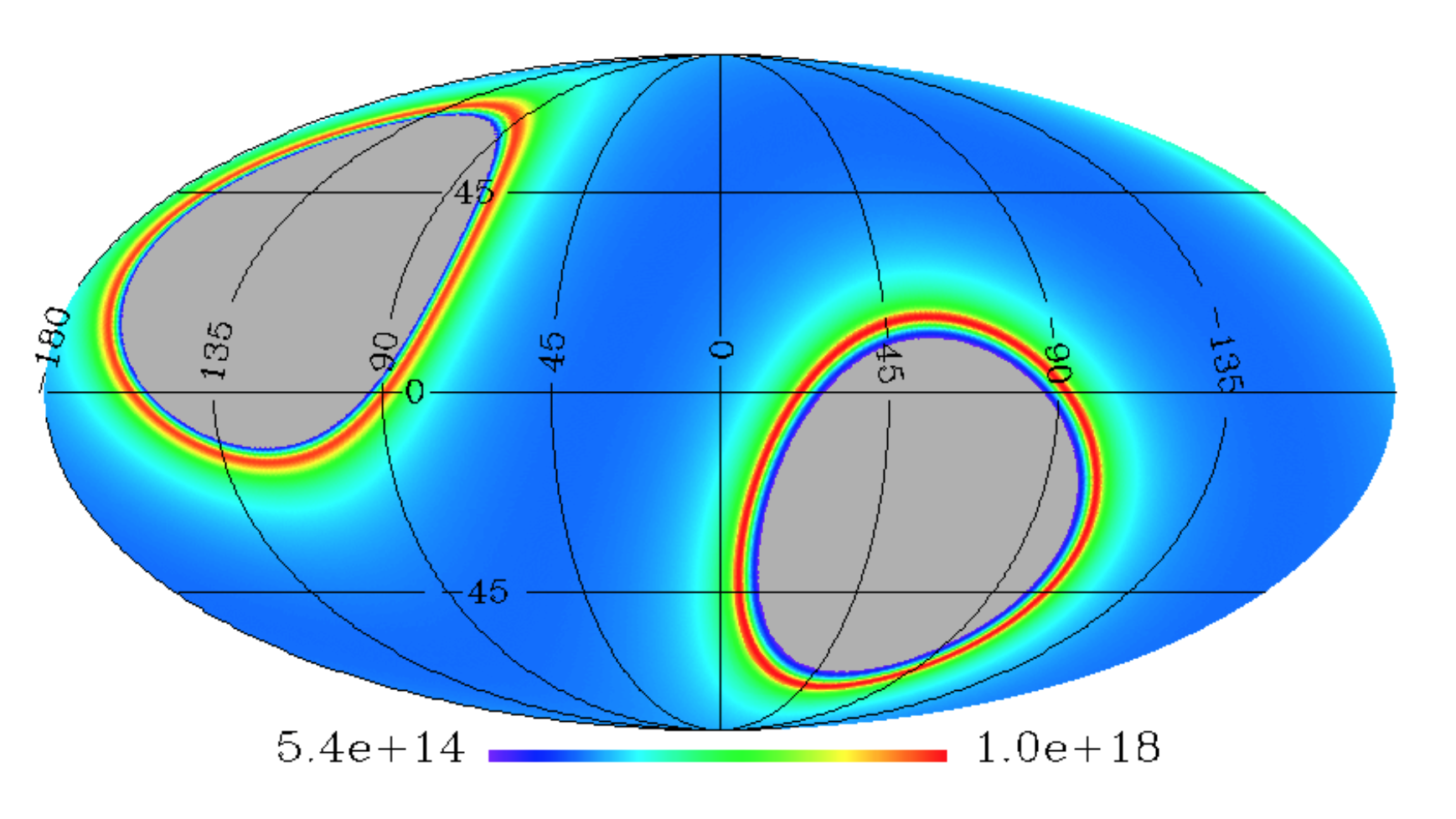,width=0.6\textwidth}  
\caption{\label{fig:exposure}
\small Three-year integrated exposure (in units of cm$^2$ s) of GRAND10k at 3~EeV in Galactic coordinates. }
\end{figure}

We present a preliminary evaluation of the GRAND sensitivity based on a simulation of the response of 90,000 antennas deployed on a square layout of 60,000 km$^2$ in a remote mountainous area (Tianshan mountains, XinJiang province, China).  The simulations are composed of three parts: I) $\nu_\tau$s interact with the rock, producing $\tau$ leptons, II) $\tau$s propagate and decay, and III) The $\tau$ decay products initiate an EAS that triggers the antennas to be read-out.  

In part I, standard rock with a density of $2.65\,\rm g\,cm^{-3}$ is assumed down to the sea level, and the Earth core is modeled following the Preliminary Reference Earth Model \cite{1981PEPI...25..297D}. We simulate the deep inelastic scattering of neutrinos with Pythia6.4, using the CTEQ5d probability distribution functions (PDF) combined with the cross sections from \cite{1998PhRvD..58i3009G}.

In part II, we simulate the propagation of the produced $\tau$s with GEANT4.9, using parameterizations of PDFs of the $\tau$ path length and proper time. We simulate $\tau$ decays with the TAUOLA package \cite{JADACH1991275}. For the above two parts, 1D tracking of a primary $\nu_\tau$ is performed down to the converted $\tau$ decay.

We have refined our previous calculation of part III \cite{GRAND2015} with numerical simulations.  We first simulate a set of $\nu_\tau$ showers with ZHaireS \cite{2012APh....35..325A} at different energies, and determine a conical volume inside which the electric field is above the expected detection threshold of the GRAND antennas.  Based on a parametrization of the shape (including angle and height) of the detection cones as a function of energy, we compute the expected shape and position of cones for each neutrino-initiated EAS in our simulation (so far in the 30-80~MHz band). We select the antennas located inside the cones, taking into account signal shadowing by mountains. When the selection consists of 8 or more neighboring antennas in direct view of the $\tau$ decay point, we count the primary $\nu_\tau$ as detected.

% Hotspot 
Our analysis finds that about 50\% of detectable neutrino interactions take place inside mountains rather than in the Earth crust. Moreover, some parts of the array provide 2-3 times higher detection rates compared to the average rate of the entire simulated area. These ``hotspots" are typically located in a low electromagnetic noise background regions and have desirable geometric features (for e.g., large mountains, slopes facing another mountain range at distances of 30-80 km). An envisioned GRAND setup is composed of smaller sub-arrays of a few 10,000 km$^2$ each, deployed solely on such favorable sites.

% sensitivity vs model predictions
Figure~\ref{fig:sensitivity} presents the differential 90\%-CL sensitivity of a GRAND simulation for a 10,000 km$^2$ hotspot and the projected full GRAND array limit after a 3-year observation. An all-flavor 90\% CL integral limit of $2\times 10^{-10}\,\rm GeV cm^{-2} s^{-1} sr^{-1}$ is derived for an unbroken $E^{-2}$ power law spectrum in an aggressive scenario (corresponding to a detection threshold of $30\,\mu \rm V\,m^{-1}$; $3.9 \times 10^{-9}\,\rm GeV cm^{-2} s^{-1} sr^{-1}$ in a conservative scenario with a detection threshold of  $100\,\mu \rm V\,m^{-1}$). For comparison, a prediction of the cosmogenic neutrino flux is shown as the grey shaded region. The flux mostly depends on three properties of UHECRs: spectral index, chemical composition, and maximum energy $E_{\rm max}$. The shaded area brackets a wide range of parameters that reflect different compositions, source evolution and Galactic to extragalactic transition models, all required to fit the UHECR spectrum measured by Auger \cite{Aab:2015bza}. 
Specifically, the upper bound corresponds to an UHECR dip model with a pure proton composition and source evolution following the star formation rate (SFR). The lower bound corresponds to a mixed composition with 30\% iron nuclei, SFR evolution and $E_{\rm max}=10^{19}\,Z$~eV.  
The grey dotted and dashed lines show the most pessimistic scenarios, where UHECRs are mostly composed of iron nuclei which barely produce neutrinos in a photo-disintegration process. We note that the best-fit models, found by a combined fit to the Auger spectrum, $X_{\rm max}$ and the standard derivation of $X_{\rm max}$ measurements \cite{2016arXiv161207155T}, correspond to a cosmogenic neutrino flux level of $\sim 10^{-9}\,\rm GeV\,cm^{-2}\,s^{-1}\,sr^{-1}$.  For comparison, we show UHE neutrino upper limits of current experiments IceCube \cite{2016PhRvL.117x1101A} and Auger \cite{PhysRevD.91.092008} (the sensitivity curve of ANITA-II \cite{2010PhRvD..82b2004G} is not shown as it would be too high for the plot). GRAND10k will be as sensitive as the proposed ARA \cite{2012APh....35..457A} and ARIANA \cite{2015APh....70...12B} experiments at 1~EeV. The full GRAND array will nominally detect even the lowest expected fluxes of cosmogenic neutrinos.

Following \cite{2011APh....34..717A}, the angular resolution of the radio array is estimated by the resolution of the time difference between pulses in different antennas. Using a 3~ns precision on the antenna trigger timing, GRAND can reach an angular resolution as accurate as 0.1$^\circ$. 

The instantaneous field of view covers a band with zenith angles between $86^\circ$ and $93^\circ$ in the simulation (the zenith angle-dependent effective area is shown in Figure~1 of \cite{GRAND2015}). The integrated exposure over 3 years of GRAND operation will cover a large fraction of the sky as shown in Figure~\ref{fig:exposure}. With stacking searches, GRAND will be sensitive to transient sources and flaring events with a duration longer than $\sim 1$~day.  Several interesting UHECR features, such as the Telescope Array hotspot and the Auger hotspot, are located close to GRAND's best sensitivity region. 

\subsection{Background rejection}
Radio background from high-energy particles other than cosmic neutrinos should be manageable. The atmospheric neutrino background is mostly negligible above $10^{16}$~eV. EAS generated by muon decay are expected at a rate of a few per century as muons have a significantly longer life time than taus. Rejecting trajectories far from the horizon should effectively suppress the EAS generated by cosmic rays.

Background event rates associated to terrestrial sources such as human activities and thunderstorms are difficult to evaluate. The background measurement of the Tianshan Radio Experiment for Neutrino Detection (TREND) \cite{2011APh....34..717A} provides a conservative estimate. TREND consisted of 50 self-triggered antennas deployed in a populated valley of the Tianshan mountains, with an antenna design and sensitivity similar to what is foreseen for GRAND.  The observed rate of events triggering six selected TREND antennas separated by $\sim$800~m over a sample period of 120 live days was found to be around 1 per day, among which two-thirds arrived as bursts of events mostly due to air plane activities. An extrapolation from this measurement leads to an expected background rate of $1\,\rm Hz$ for GRAND for a trigger algorithm based on coincident pulses on neighboring antennas and a rejection of event bursts.

Specific amplitude patterns on the ground (such as beamed emission along the shower axis and signal enhancement along the Cherenkov ring \cite{2012PhRvD..86l3007A}), and polarization of signals \cite{2014PhRvD..89e2002A} are strong indicators of neutrino-initiated EAS. We are investigating the possibilities of using these signatures as efficient background discrimination tools through both simulations and experiments. 
%The GRANDproto project \cite{GRANDproto2017}, which is a hybrid detector composed of 35 3-arm antennas (which enable us to reconstruct the wave polarization) and 24 scintillators, will cross-check radio-events selected using the polarization signature. 
 
\section{Science case}
\begin{figure}[h]
\vspace*{-9mm}
\centering
\epsfig{file=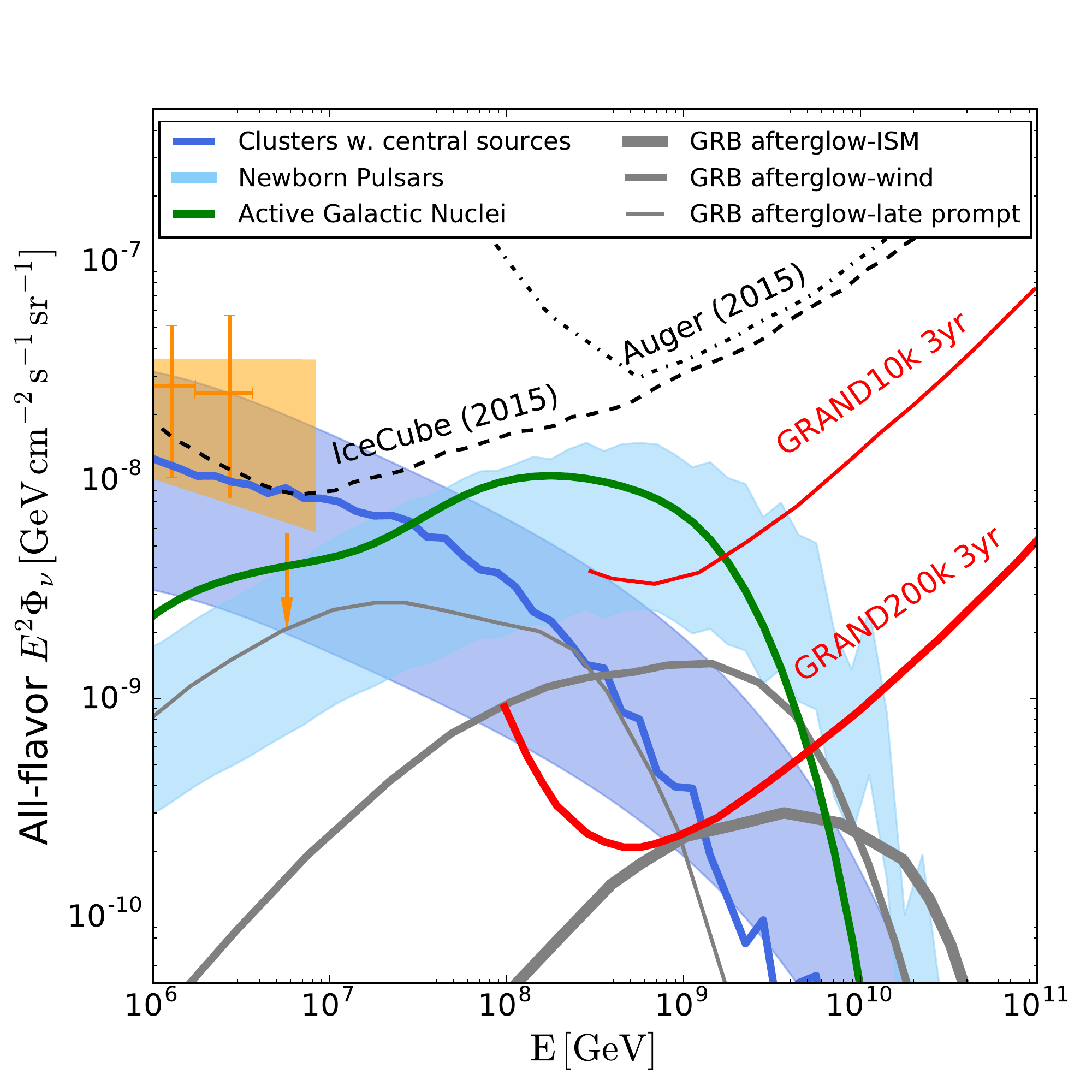,width=0.65\textwidth}  
\caption{\label{fig:source}
{\small Three-year differential sensitivity of the GRAND10k and GRAND full array comparing to fluxes of neutrinos from interactions inside astrophysical sources. Four source classes that can explain UHECR measurements are shown: galaxy clusters with central sources \cite{2008ApJ...689L.105M, 2017arXiv170400015F}, fast-spinning newborn pulsars \cite{2014PhRvD..90j3005F}, active galactic nuclei \cite{2015arXiv151101590M}, and afterglows of gamma-ray bursts \cite{PhysRevD.76.123001}. 
High-energy neutrino observed by IceCube \cite{2015arXiv151005223T}, UHE neutrino sensitivities of IceCube \cite{2016PhRvL.117x1101A} and Auger \cite{PhysRevD.91.092008} are shown for reference.} }
\end{figure}

%  source neutrinos 
In addition to a guaranteed detection of cosmogenic neutrinos, GRAND will allow a differentiation of the source types by measuring neutrinos produced inside sources. In Figure~\ref{fig:source}, we show fluxes of neutrinos from four typical classes of astrophysical sources that can explain UHECR measurements, including galaxy clusters with central sources \cite{2008ApJ...689L.105M, 2017arXiv170400015F}, fast-spinning newborn pulsars \cite{2014PhRvD..90j3005F}, active galactic nuclei \cite{2015arXiv151101590M}, and afterglows of gamma-ray bursts \cite{PhysRevD.76.123001}. The GRAND sensitivity is sufficient to observe neutrinos from sources in all cases. 

% identify individual sources 
A $5\sigma$ identification of individual point sources out of a diffuse background requires $\sim 100-1000$ events with a sub-degree angular resolution for sources that have a local density of $10^{-9}-10^{-7}\,\rm Mpc^{-3}$ \cite{FKMMO16}. Assuming a neutrino flux of $10^{-8}\,\rm GeV\,cm^{-2}\,s^{-1}\,sr^{-1}$, the GRAND neutrino sensitivity corresponds to a detection of $\sim 100$ events after three years of observation. The unprecedented sensitivity of GRAND is crucial for a direct identification of UHE neutrino point sources. 

% UHECRs 
GRAND will observe UHECRs with an effective area that is at least an order of magnitude larger than Auger. The high statistics will resolve any small-scale anisotropies and features near the end of the cosmic ray spectrum.
% It also allows rich particle physics study at the highest energies with for example nuclei-air cross section measurements well above LHC energies. 
GRAND will also reach an UHE photon sensitivity exceeding that of current experiments. Future simulation studies will be dedicated to performance of energy and $X_{\rm max}$ reconstructions. 

% other sciences FRBs 
GRAND will contribute in a unique way to the measurement of fast radio bursts (FRB) and giant radio pulses by collecting unprecedented statistics at low frequencies \cite{2016sf2a.conf..347Z}. Simulations are being performed to quantify the detection power with incoherent sum of FRB spectra computed at each antenna.   
%For instance, GRAND could record giant pulses from the Crab pulsar above 5 Jy at 200 MHz with a rate about 200 per day. 
The possibility of GRAND's contribution to additional science topics, such as the study of the epoch of reionization, is also being investigated. 

\section{Development plan}  
The GRAND project aims at building a next-generation neutrino telescope composed of a radio antenna array deployed over 200,000 km$^2$. Preliminary simulations indicate that 5\% of the array deployed in favorable sites will improve the sensitivity over that of current-generation telescopes by an order of magnitude. The full array will offer a sensitivity that ensures a detection of cosmogenic neutrinos in the most pessimistic scenario, and an identification of individual point sources in an optimistic scenario. GRAND will also be a powerful instrument for UHECR observations with high statistics. Simulation and experimental work is ongoing on technological development and background rejection strategies. 

The GRAND development plan consists of several steps.  Presently, deployment of GRANDproto \cite{GRANDproto2017}, which is a hybrid detector composed of 35 3-arm antennas and 24 scintillators, is underway at the site of the TREND experiment, in the Tianshan mountains. The GRANDproto project will demonstrate that EAS can be detected with an autonomous radio array with both high efficiency and high purity. 
Following this step, a dedicated setup with a size of 300 km$^2$ will be deployed.  This array will establish the autonomous radio detection of very inclined EAS with  excellent background rejection, as a validation for the future GRAND layout. 

\acknowledgments
\small {The GRAND and GRANDproto project are supported by the France China Particle Physics Laboratory, the Institut Lagrange de Paris, the APACHE grant (ANR-16-CE31- 0001) of the French Agence Nationale de la Recherche, the Natural Science Foundation of China (Nos.11135010, 11105156, 11375209 and 11405180), the Chinese Ministry of Science and Technology, the Youth Innovation Promotion Association of Chinese Academy of  Sciences, and the São Paulo Research Foundation FAPESP (grant 2015/15735-1).}

\bibliographystyle{ieeetr}
\bibliography{GRAND}

%\begin{thebibliography}{99}
%\bibitem{...}
%....
%\end{thebibliography}

\end{document}